# Peer Review and the Diffusion of Ideas

Binglu Wang[1,2,3], Zhengnan Ma[4], Dashun Wang[1,2,3,4], Brian Uzzi[1,3,4†]

[1] Kellogg School of Management, Northwestern University

[2] Center for Science of Science and Innovation, Northwestern University, Evanston, IL, USA

[3] Northwestern University Institute on Complex Systems, Evanston, Illinois

[4] McCormick School of Engineering, Northwestern University, Evanston, IL, USA

[†] *Correspondence should be addressed to B.U. ([uzzi@kellogg.northwestern.edu](uzzi@kellogg.northwestern.edu))*

**Abstract**

This study examines a fundamental yet overlooked function of peer review: its role in exposing reviewers to new and unexpected ideas. Leveraging a natural experiment involving over half a million peer review invitations covering both accepted and rejected manuscripts, and integrating high-scale bibliographic and editorial records for 37,279 submitting authors, we find that exposure to a manuscript's core ideas significantly influences the future referencing behavior and knowledge of reviewer invitees who decline the review invite. Specifically, declining reviewer invitees who could view concise summaries of the manuscript's core ideas not only increase their citations to the manuscript itself but also demonstrate expanded breadth, depth, diversity, and prominence of citations to the submitting author's broader body of work. Overall, these results suggest peer review substantially influences the spread of scientific knowledge. Ironically, while the massive scale of peer review, entailing millions of reviews annually, often drives policy debates about its costs and burdens, our findings demonstrate that precisely because of this scale, peer review serves as a powerful yet previously unrecognized engine for idea diffusion, which is central to scientific advances and scholarly communication.

**Introduction**

Peer review is a cornerstone of scientific practice, but its origins reveal an often-overlooked function beyond the familiar roles of providing feedback and quality control. Originally formalized in the 19th century by journals such as *Philosophical Transactions*, peer review was introduced to entice scientists away from informal socialization of their discoveries and towards formal dissemination through scientific journals [1]. The promise was simple but powerful: by submitting work to a journal, scientists were assured a readership—at least the handful of peers who would review their submissions. Thus, peer review emerged not merely as science's gatekeeping tool, but fundamentally as an engine of knowledge diffusion, promoting the spread and adoption of new ideas.

Over the ensuing centuries, peer review's foundational role as a diffusion mechanism has been overshadowed by its critical function in maintaining research integrity and quality [2-4]. Today, individual scientists collectively spend about 100 million hours per year on peer review [5], roughly equivalent to the time it would take to write millions of grant proposals [6], yet the system's efficiency and effectiveness have remained subjects of ongoing debate and improvement efforts [7-9].

Despite the extensive focus on its evaluative role, little attention has been paid to peer review's foundational role as a mechanism of diffusing new ideas, despite the fact that exposure and access to new ideas support innovation in science as much as the filtering process [10-12]. Yet, even as the overall volume of scientific knowledge has grown rapidly [13, 14], knowledge search progressively emphasizes a small fraction of work published in the leading journals, informatics search tools present knowledge by popularity, and lengthening journal publication lags slow access to new knowledge [13, 15, 16]. Within this context, peer review invitations represent unique micro-exposures to emerging research, potentially counteracting trends towards narrowing scholarly attention by serendipitously exposing reviewer invitees to new ideas.

Here, we quantify peer review's original promise as a conduit of knowledge diffusion. For example, does early exposure to research work during peer review promote the adoption of new ideas among reviewer invitees? Do the characteristics of reviewer invitees' citations to work they are exposed to in the review invitations expand the breadth, depth, diversity, and prominence of citations, enriching science and spreading peer review costs over more benefits? These questions have remained elusive due to substantial empirical challenges. Such research requires comprehensive, longitudinal data on editorial decisions for both accepted and rejected manuscripts, detailed professional identities of authors and reviewer invitees, citation behaviors, and insights into the nuances of editorial processes—data typically restricted or inaccessible, partly due to the inherent anonymity and confidentiality that underpins peer review.

Here, we conduct one of the first studies of links between peer review and the diffusion of new and related ideas. We use a unique 10-year, longitudinal dataset that combines 551K peer review invitations from a leading medical journal with detailed bibliometric records on the professional identities of authors and reviewer invitees. We then merge this data with SciSciNet [17] and Web of Science (WoS), large-scale publication and bibliographic databases. Together, the data allow us to identify the reviewer invitees' history of research, expertise, response to the review, and what they later cite and publish, including new and related work by the submitting author. This rare combination of editorial records and bibliographic metadata enables the first systematic test of whether peer review is related to the spread of scientific ideas.

We quantify a reviewer invitees' adoption of new and related ideas into their research by examining their citation behavior toward the submitting author's work before and after exposure to the review invitations, employing a difference-in-differences (DiD) regression framework. In addition to overall citation frequency, we quantify citation quality through measures of prominence, as well as citation breadth, depth, and diversity.

A key analytical challenge in quantifying peer review's diffusion effect lies in reviewer self-selection. Comparing individuals who actively reviewed manuscripts with those who did not can be problematic because reviewers typically choose manuscripts that align closely with their interests, expertise, and prior knowledge, thus introducing significant selection bias. Our methodological innovation addresses this challenge by focusing exclusively on individuals who did not conduct the review. Specifically, we compare two distinct groups of invited reviewers: those who actively decline the invitation by clicking a link (Click-Decline) and those passively declined by the journal's automated system (Auto-Decline). While these two groups are comparable in that neither group engages fully with the manuscript, they differ in that the Click-Decline reviewer invitees likely viewed essential manuscript metadata, such as title, abstract, and authorship before declining, whereas Auto-Decline reviewer invitees likely did not. This approach allows us to isolate incidental exposure to a manuscript's core ideas independent of self-selection. To strengthen our assumption about differences in Click-Decline and Auto-Decline groups' likely exposure to a manuscript's core ideas, we further exploit a natural experiment: the quasi-random timing of peer review invitations—sent on either a workday or weekend—which affects the likelihood of invitee engagement independently of their research interests or citation intentions. [18].

Across diverse tests and robustness checks, we consistently find that concise exposure through peer review invitations meaningfully promotes the dissemination of new ideas. These findings demonstrate that peer review functions not only as a gatekeeping institution but also as a subtle yet powerful engine of knowledge diffusion – an essential and previously overlooked benefit embedded within this foundational institution of science.

**Data**

We analyzed the complete peer review records of one journal from a premier publisher of medical research over a ten-year span (2012-2022), encompassing 37K submissions and 551K review records[1]. Each record captures the full invitation process, including metadata about the manuscript—title, abstract, author names, and the corresponding author—as well as timestamped reviewer invitees' responses (Figure 1A). The data were collected by the journal through their ScholarOne intake platform (formerly Manuscript Central). Besides the fine-grained data collected by ScholarOne, a unique feature of these data is that ScholarOne's broad use across science standardizes scholarly editorial workflows and peer review processes despite journal-specific or author-specific characteristics of different fields. Over 9,000 diverse journals and top publishers use ScholarOne (www.silverchair.com) to manage over three million manuscripts annually as part of their peer review and submission processes. Thus, while peer review's filtering role can be moderated by journal-specific and author-specific characteristics [4],

---

[1] This study was IRB approved (STU00219321).

ScholarOne employs the same standardized intake and review system similarly across journals, manuscripts, and reviewer invitees.

Upon receiving an invitation to review, reviewer invitees may respond in one of three ways (Figure 1B): (1) Accept—clicking to accept the invitation; (2) Click-Decline—explicitly declining by clicking "Declined" or "Unavailable"; or (3) Auto-Decline—not responding to the invitation, with the system logging it as a passive decline.

This setting offers a rare window into an otherwise invisible part of scientific communication: micro-exposures to emerging scientific content—delivered through manuscript titles, abstracts, and author lists—well before publication. The data's fine-grained structure allows us to observe and differentiate these exposures at scale.

We then link this dataset to SciSciNet [17] and WoS, allowing us to track individual-level bibliometric records and the subsequent citation behaviors of each reviewer invitee over time (See details in SI S1). For every reviewer invitee, we observe both the manuscripts they were invited to evaluate and their longitudinal publication and citation trajectories.

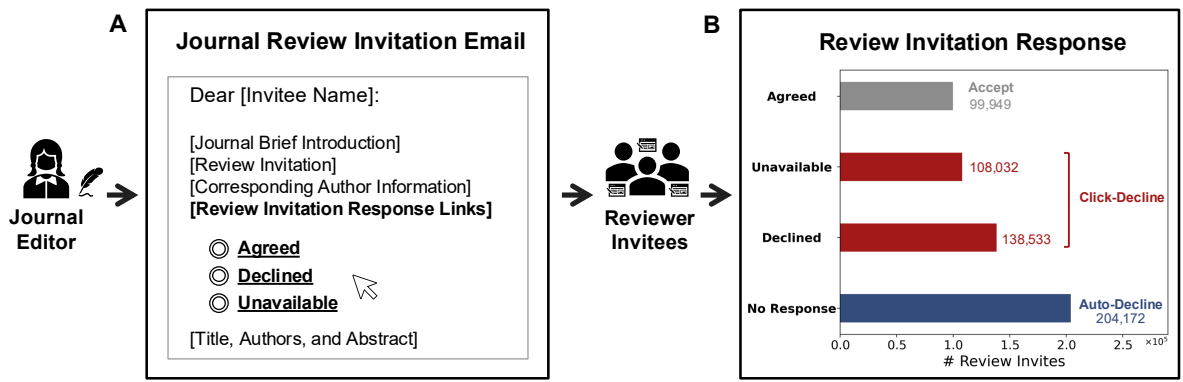

**Figure 1. The peer review invitation process and classification of reviewer invitees' responses.** (A) Editors send email invitations containing metadata about the journal and manuscript, along with embedded response links: "Agreed," "Declined," and "Unavailable." (B) We group reviewer invitees based on their responses to the invitation: the Click-Decline group includes 247K invitees (44.77%) who actively decline the review request; the Auto-Decline group consists of 204K (37.08%) invitees who passively non-respond to the review request; and the Accept group has 100K individuals (18.15%) who actively accept the review.

**Results**

Figure 2A presents the average annual citation trajectories of three reviewer invitee groups: Accept (gray), Click-Decline (red), and Auto-Decline (blue). Prior to the review invitation, citation patterns for the Click-Decline and Auto-Decline groups are nearly identical (Wald test: $p = 0.543$), supporting the comparability of these two non-reviewing groups. In contrast, the Accept group displays a markedly higher baseline citation rate to the corresponding author, vividly illustrating the self-selection challenge: reviewers who agree to evaluate a manuscript are systematically more familiar with or interested in the authors' work, biasing post-review comparisons. Consequently, to ensure valid causal inference, our analysis strategically focuses on comparing the Click-Decline and Auto-Decline groups, thereby isolating incidental exposure effects from reviewer self-selection.

While the citation trajectories for the Click-Decline and Auto-Decline groups were indistinguishable prior to the review invitation, they diverge notably afterward (Figure 2A). Reviewer invitees who actively clicked to decline—thus likely exposed to the manuscript's metadata—show a sustained increase in citations to the corresponding author's work. In contrast, the Auto-Decline group, who passively ignored the invitation, maintains a flat trajectory. Five years after the invitation, Click-Decline reviewer invitees cite the corresponding author's work 29.09% more than the Auto-Decline group. These results suggest that concise exposure to a manuscript's core ideas during the invitation process can meaningfully and persistently influence the diffusion of scientific knowledge, despite reviewer invitees not accessing the full manuscript during the process.

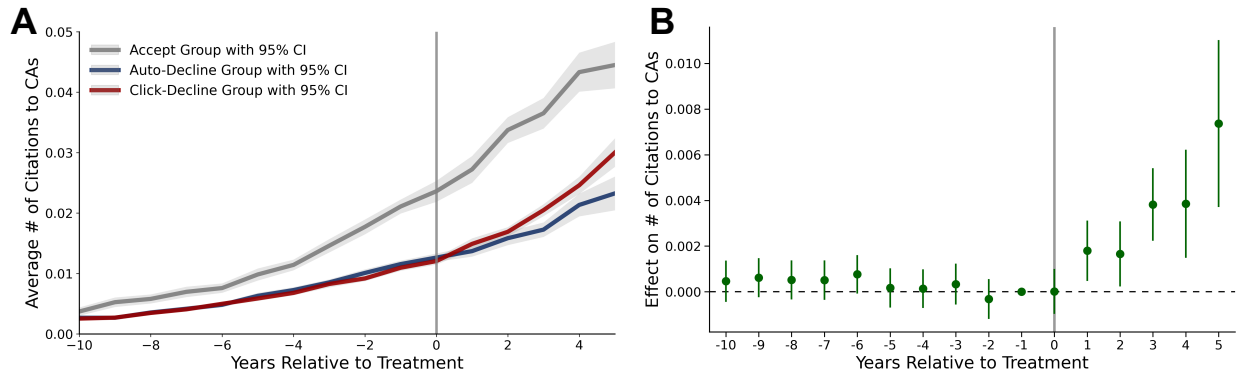

**Figure 2. Citations to corresponding authors over time and event-study estimates. (A)** Average number of citations from reviewer invitees to the corresponding authors (CAs) of manuscripts they were invited to review, plotted by year relative to the invitation (Year 0). The plot compares the Accept group, Click-Decline (treatment) group and Auto-Decline (control) group. Shaded areas represent 95% confidence intervals. **(B)** Event-study estimates from the regression, restricted to the Click-Decline and Auto-Decline groups. The model includes relative-year indicators interacted with treatment status to estimate the dynamic effect of minimal exposure on subsequent citations to the corresponding authors.

To systematically test the observed effect controlling for moderators [19, 20], we employ a DiD regression model on the Click-Decline and Auto-Decline subsample:

$$y_{it} = \beta_0\, Treat_i \times Post_t + \beta_1\, Treat_i + \beta_2\, Post_t + \beta_3\, C_i + \epsilon_{it}$$

where $y_{it}$ denotes the number of citations from reviewer invitee $i$ to the corresponding author in year $t$; $Treat_i$ is an indicator for membership in the Click-Decline group; and $Post_t$ marks years after the invitation. The interaction term, $Treat_i \times Post_t$, captures the differential change in citations for the treatment group post-invitation. Figure 2B reports event-study estimates, confirming parallel pre-trends and revealing a statistically significant post-invitation increase in citation behavior for the Click-Decline group.

Model (1) in Table 1 reports the baseline model estimates: Click-Decline group cites corresponding authors 0.0021 more times than Auto-Decline group after the invitation ($p = 0.000$). While modest in absolute terms, this reflects a 14% relative increase compared to the Auto-Decline group's post-invitation citation rate. Event study estimates reveal that the increase begins in the invitation year and persists over time. To account for potential confounders, we extend the DiD model with covariates in Model (2) capturing: (1) author–reviewer proximity (prior co-authorships and mutual citations), to address pre-existing intellectual connections before invitation; (2) reviewer invitees' experience (publication count and $h$-index), to control for seniority and scholarly productivity before invitation; and (3) engagement with the journal (past publications in and citations to the journal), to capture alignment with the journal's content before invitation. See statistical descriptions of these variables in SI S2.

Including these controls modestly changed the treatment effect to 0.0018 ($p = 0.000$), corresponding to a 12% relative increase. This reinforces that the observed differences stem from exposure to the invitation rather than from individual-specific characteristics. We further validate our findings using Two-Way Fixed Effects (TWFE) and Staggered DiD models [21], which address treatment timing heterogeneity and potential confounding from staggered exposures. These specifications yield consistent estimates, detailed in SI S3, confirming the robustness of the effect.

These findings support the hypothesis that peer review facilitates the diffusion of new and related scientific ideas. While exposure in this setting typically involves only a brief inspection of a manuscript's title and abstract, the magnitude and persistence of its effect become substantial when aggregated across the millions of peer review invitations extended annually.

TABLE 1: BASELINE ESTIMATES

|  | (1) | (2) |
|---|---|---|
| $Treat_i \times Post_t$ | 0.0021*** | 0.0018*** |
|  | (0.0004) | (0.0004) |
| $Treat_i$ | -0.0003* | 0.0003* |
|  | (0.0002) | (0.0001) |
| $Post_t$ | 0.0087*** | 0.0079*** |
|  | (0.0003) | (0.0003) |
| # Collaboration for Both |  | -0.0059 |
|  |  | (0.0037) |
| # Citation Reviewer to Author |  | 0.0060*** |
|  |  | (0.0004) |
| # Citation Author to Reviewer |  | -0.0004* |
|  |  | (0.0002) |
| Reviewer # Publications (log) |  | -0.0035*** |
|  |  | (0.0001) |
| Reviewer H-index (log) |  | 0.0048*** |
|  |  | (0.0002) |
| Reviewer # Pub on Journal |  | -0.0001** |
|  |  | (0.0000) |
| Reviewer # Ref to Journal |  | -0.0000* |
|  |  | (0.0000) |
| Const | 0.0065*** | 0.0079*** |
|  | (0.0001) | (0.0008) |
| Observations | 4,755,571 | 4,755,571 |
| No. clusters | 450,737 | 450,737 |
| R-squared | 0.0014 | 0.1272 |

**Table 1. The effect of exposure via review invitation on downstream citations to corresponding authors.** This table presents regression estimates using a panel dataset of reviewer–manuscript pairs spanning ten years before to five years after the review invitation. The dependent variable is the average annual number of citations made by the reviewer invitees to the corresponding author. Model (1) shows that reviewer invitees who Click-Declined the invitation cite corresponding authors 0.0021 more times per year after the invitation than those who Auto-Declined ($p = 0.000$). This effect represents a 14% relative increase compared to the post-invitation citation rate of the Auto-Decline group. Model (2) adds covariates controlling for prior connections to the corresponding author (co-authorships and mutual citations), reviewer invitees' experience (career H-index, total publications), and prior engagement with the journal. The estimated effect remains statistically significant at 0.0018 citations per year ($p = 0.000$), reflecting a 12% relative increase. Robust standard errors clustered at the reviewer invitee level are reported in parentheses (*** $p < 0.01$, ** $p < 0.05$, * $p < 0.1$).

## Causal Identification

To further strengthen the causal interpretation of our findings, we implement an instrumental variable approach using a natural experiment that induces exogenous variation in exposure to peer review invitations [18, 22]. Specifically, we exploit the quasi-random variation in the timing of invitation emails—comparing those sent on weekdays (Monday through Friday) to those sent on weekends (Saturday and Sunday). Of the 551K total invitations in our dataset, 90.06% (496K) were issued on weekdays, whereas 9.94% (55K) were sent on weekends.

This timing-based variation serves as a plausibly exogenous shock to exposure likelihood. Given that scientists systematically exhibit reduced email activity during weekends, invitations sent on weekends experience significantly lower response rates. Consequently, weekend timing serves as a natural instrument for exposure to manuscript content. Indeed, because the day on which ScholarOne dispatches a review invitation is unrelated to reviewer-specific factors such as research interests, prior familiarity with the manuscript, or future citation intentions, our instrumental variable approach satisfies the exclusion restriction, as illustrated conceptually in Figure 3.

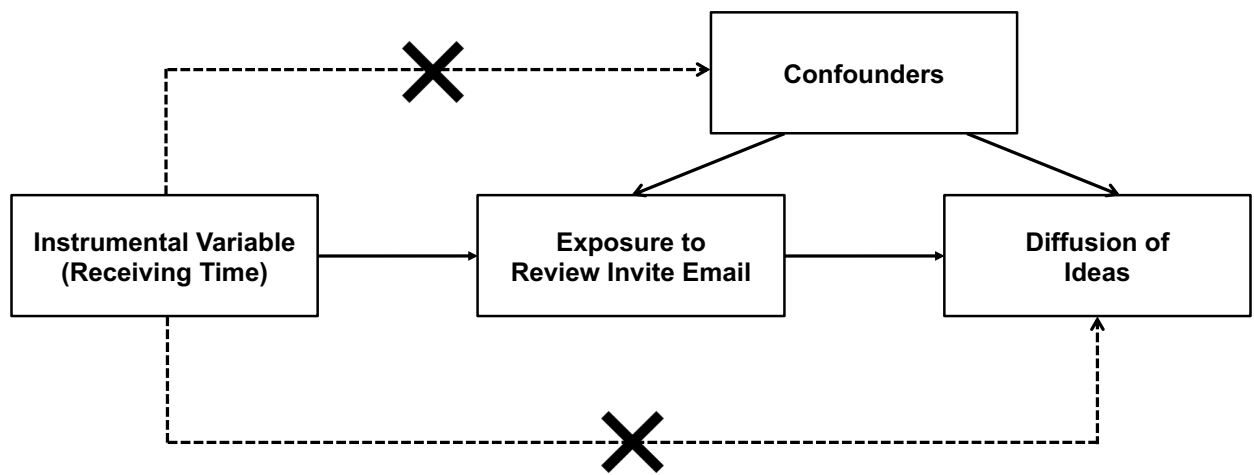

**Figure 3. Instrument Variable Framework.** The diagram illustrates the identification strategy: the instrument influences adoption of related ideas solely through its effect on exposure to the review invitation. It is assumed to be independent of unobserved confounders and to have no direct effect on the outcome, as denoted by the dashed arrows crossed out. This structure enables causal inference by isolating exogenous variation in exposure.

We formalize this identification strategy using a two-stage least squares (2SLS) regression model:

$$1^{st} \text{ Stage: } \overline{Treat_i \times post_t} = \beta_0 IV\_weekend_i + \varepsilon_{it}$$
$$2^{nd} \text{ Stage: } y_{it} = \beta_1 + \beta_2 \overline{Treat_i \times post_t} + \varepsilon_{it}$$

Here, $\overline{Treat_i \times Post_t}$ captures exposure to the invitation during the post-invitation period, and $IV\_weekend_i$ is a binary instrument indicating whether the invitation was sent on a weekend. The unit of analysis remains the reviewer–manuscript pair.

The first-stage regression yields an F-statistic of 1275.8, well above conventional thresholds for weak instruments [23], confirming that weekend timing is a strong predictor of reduced exposure. Full first-stage estimates are reported in SI S4.

Table 2 presents the second-stage estimates from the 2SLS model. In the baseline specification without additional covariates (Model 1), the estimated treatment effect is 0.0288 ($p = 0.005$), meaning that reviewer invitees exposed to the invitation content cited the corresponding author 0.0288 more times post-invitation than their unexposed counterparts. After including the full set of covariates in Model (2) — capturing prior author–reviewer connections, reviewer invitees' characteristics, and journal engagement — the estimated treatment effect remains statistically significant and decreases slightly to 0.0213 ($p = 0.027$). These estimates are robust across specifications and validate our main conclusion: incidental exposure to new research through peer review invitation emails can causally influence subsequent scholarly behavior.

Notably, the 2SLS estimates exceed the baseline difference-in-differences models, suggesting that unobserved confounding likely attenuated earlier results. By addressing endogeneity in exposure, the IV approach offers a more credible estimate of the true causal effect. Together, these findings reinforce the conclusion that peer review is a pervasive conduit through which new and related scientific ideas are adopted not necessarily by the reviewer invitees who come to know the work deeply through formally reviewing a paper, but by reviewer invitees who in the process of declining a review are likely exposed to a concise summary of new ideas.

TABLE 2: INSTRUMENTAL VARIABLE 2SLS

|  | (1) | (2) |
|---|---|---|
| $\overline{Treat_\iota} \times Post_t$ | 0.0288*** | 0.0213** |
|  | (0.0104) | (0.0096) |
| # Collaboration for Both |  | -0.0044 |
|  |  | (0.0042) |
| # Citation Reviewer to Author |  | 0.0061*** |
|  |  | (0.0005) |
| # Citation Author to Reviewer |  | -0.0003 |
|  |  | (0.0002) |
| Reviewer # Publications (log) |  | -0.0039*** |
|  |  | (0.0002) |
| Reviewer H-index (log) |  | 0.0063*** |
|  |  | (0.0003) |
| Reviewer # Pub on Journal |  | -0.0001** |
|  |  | (0.0001) |
| Reviewer # Ref to Journal |  | -0.0000* |
|  |  | (0.0000) |
| Const | 0.0042* | 0.0050** |
|  | (0.0024) | (0.0021) |
| Observations | 742,086 | 742,086 |
| No. clusters | 450,737 | 450,737 |
| R-squared | 0.0033 | 0.1641 |

**Table 2. Instrumental variable estimates of the causal effect of exposure via review invitations on downstream citations.** This table reports second-stage estimates from a two-stage least squares (2SLS) regression model. The outcome is the same as Table 1, average annual number of citations from the reviewer invitees to the corresponding author. Exposure is instrumented using a binary indicator for whether the invitation was sent on a weekend, exploiting quasi-random variation in reviewer invitees' email engagement. Model (1) presents the baseline 2SLS estimate without additional covariates: reviewer invitees exposed to invitation metadata cite the corresponding author 0.0288 more times per year ($p = 0.005$) than unexposed reviewer invitees. Model (2) includes prior author–reviewer ties, reviewer invitees' characteristics, and journal engagement history; the estimated effect remains statistically significant at 0.0213 ($p = 0.027$). The first-stage F-statistic is 1275.8 for Model (1) and 1220.48 for Model (2), far exceeding conventional thresholds for weak instruments, supporting the strength of the weekend-sent instrument. These results are robust to model specification and exceed the corresponding difference-in-differences estimates, suggesting that our baseline effects are conservatively estimated. Robust standard errors clustered at the reviewer invitee level are reported in parentheses (*** $p < 0.01$, ** $p < 0.05$, * $p < 0.1$).

**Attention, Breadth, Depth, Diversity, and Prominence of Citation Behavior Change**

We extend our analysis to six complementary dimensions of post-invitation scholarly behavior, each capturing distinct facets of how reviewer invitees may engage with the corresponding author's work post-exposure. Direct Attention captures whether reviewer invitees later cite the specific manuscript they were invited to review, conditional on that manuscript being published. Breadth refers to whether reviewer invitees cite any work by the corresponding author beyond the focal manuscript, indicating a broader engagement with the author's research program. Depth (as measured in our previous analysis) evaluates how extensively the reviewer invitees cite the corresponding author's work within their own publications; higher values suggest more intensive engagement. To account for varying reference list lengths, we also construct a Normalized Depth measure, which adjusts for the total number of references per publication, thus capturing the relative salience of the author's work within each citing paper. Diversity quantifies how many distinct papers by the corresponding author are cited, providing a measure of the range of intellectual influence. Finally, Prominence assesses the prominence of the corresponding author's work within the reviewer invitee's reference list, operationalized using the reciprocal of the earliest citation position in papers as a proxy for conceptual centrality. To measure this, we use the WoS database to extract the citation position within each paper's reference list. See SI S5 for detailed variable definitions.

Figure 4 presents comparisons across six distinct dimensions of reviewer invitees' citation behavior, contrasting the Click-Decline and Auto-Decline groups in the period following the invitation using two-sample t-tests. Group means are shown with 95% confidence intervals, with significance levels annotated. We find that, after the invitation, the Click-Decline group exhibits statistically significant increases across all six dimensions, indicating that peer review promotes not only the diffusion of new and related ideas but also enhances the breadth, depth, diversity, and prominence, with which these ideas are engaged and incorporated into a scholar's research. Corresponding pre-invitation comparisons are presented in SI S5 (Figure S3), confirming the robustness and conservative nature of these results. We further confirm these results using a difference-in-differences (DiD) regression model on the same outcome variables. The estimated treatment effects remain robust and consistent across all six dimensions (SI S5 Table S6).

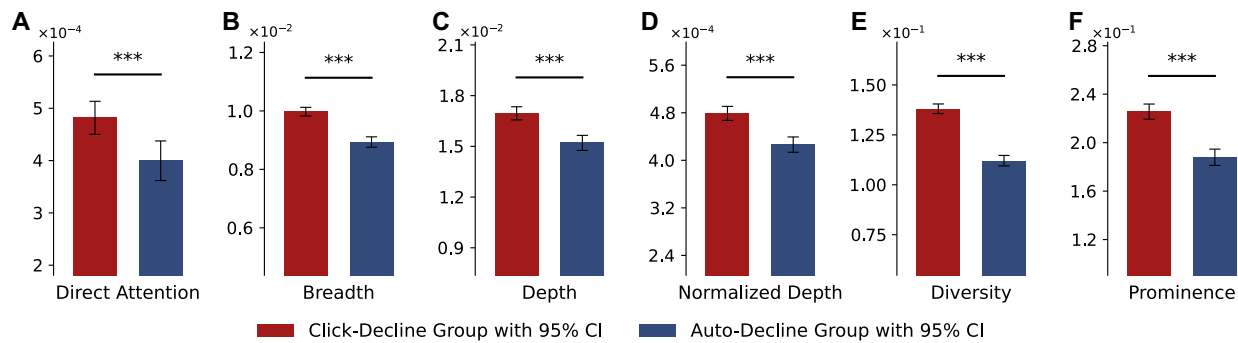

**Figure 4. Post-invitation comparisons across six dimensions.** We compare the Click-Decline and Auto-Decline groups using two-sample t-tests on six citation-based outcome measures for post-invitation periods. Group means are displayed with 95% confidence intervals, with significance levels and conceptual groupings annotated. Outcome variables include: (A) Direct Attention (proportion of reviewer invitees' publications citing the invited manuscript); (B) Breadth (proportion of reviewer invitees' publications citing any work by the corresponding author); (C) Depth (average number of references to the corresponding author); (D) Normalized Depth (share of references per paper attributed to the author); (E) Diversity (number of unique papers by the author cited); and (F) Prominence (the reciprocal of the earliest reference position of the author's work in reviewer invitees' papers). Color coding indicates conceptual clusters. Following the invitation, Click-Decline invitees subsequently cite the corresponding author with significantly greater direct attention, breadth, depth, diversity, and prominence than Auto-Decline invitees. (∗∗∗ $p < 0.01$, ∗∗ $p < 0.05$, ∗ $p < 0.1$).

## Discussion

Our findings demonstrate that peer review is not merely a quality control institution but a powerful, subtle driver of scientific idea diffusion—a previously unrecognized role embedded within this foundational institution of science. Using a novel comparison of click-decline and auto-decline groups, we find that incidental exposure to concise manuscript summaries significantly and durably influences the breadth, depth, diversity and prominence of reviewer invitees' subsequent citations. Given the enormous scale of peer review, with millions of invitations annually, this uncovered effect likely constitutes a substantial yet overlooked mechanism shaping the dissemination and integration of scientific knowledge.

Theoretically, our results broaden existing understanding of how researchers acquire new knowledge. Prior work has primarily emphasized interactive, relationship-oriented channels, such as teams, mentorship, seminars, networks, or leadership [11, 19, 24-27]. Our findings introduce an important complementary dimension: concise, asynchronous exposure to novel ideas through peer review invitations. This channel is particularly valuable in a scientific environment increasingly dominated by informatics-mediated search—which presents popular rather than new knowledge first—and traditional journal publishing, which can slow the presentation of new ideas due to the lengthening of review times and publication lags.

Further, this work speaks to the broader policy questions of how learning takes place in science's increasingly information-rich environment. What is the right combination of deep relationship-oriented learning and concise learning? Does the combination potentially change over a career, team demographics, mentors, fast-moving fields (e.g., AI) vs. slow-moving fields (e.g., math), vanguard researchers that push the boundaries of fields, or among scholars pivoting to new research areas [14, 19, 27, 28]. Moreover, our findings raise novel considerations for journal editors and publishers: as editors select reviewers—an increasingly scrutinized process aimed at fostering diversity and inclusivity—they may also implicitly shape where and how scientific knowledge flows. This underscores an important yet underappreciated editorial responsibility, suggesting that strategic reviewer selection could influence not only manuscript quality assessment but also the broader diffusion and integration of new ideas across scientific communities. Lastly, LLMs are a promising addition to the review process and could provide new technological avenues for presenting knowledge in new and useful ways [29].

Several limitations merit consideration. While we have carefully identified the effect of peer review on idea diffusion, questions remain about how concisely summarized information is most effectively communicated. With research increasingly demonstrating that the semantic underpinnings of papers and grants work hand-in-hand with a paper's robust statistics in showing the merits of good ideas to reviewer invitees [30], it was beyond this study's scope to examine how variation in the presentation of concisely summarized new ideas may impact reviewer invitees' reactions [31]. Related to the linguistic nature of science, while our natural experiment was high-scale and based on a standardized science-wide peer-review system that suggests that the findings are extrapolatable to other journals, future research may further confirm and generalize our findings by involving diverse journals and disciplines.

At a time when peer review faces increasing scrutiny for inefficiencies, high costs, and reviewer fatigue, our findings highlight significant hidden benefits. Peer review emerges as a unique and scalable conduit for idea diffusion, stimulating unexpected intellectual engagement and enriching scholarly discourse. Ironically, although the enormous volume of peer reviews conducted annually is often seen primarily as a burden, our results reveal that precisely because of this volume, peer review quietly yet substantially shapes the trajectory of scientific progress. This previously unrecognized mechanism underscores peer review's enduring and essential contribution—not merely as science's gatekeeper, but also as an influential force driving innovation and the advancement of diverse ideas.

**Data Availability**

The SciSciNet data are publicly available at https://doi.org/10.6084/m9.figshare.c.6076908.v1. The Web of Science (WoS) data were used under license and are not publicly available. Peer review data contain sensitive information and cannot be shared publicly. The data necessary to reproduce the main figures in this study will be made available by the authors upon reasonable request.

**Code Availability**

The code used to generate the main figures and perform statistical analyses will be made available by the authors upon reasonable request.

**Acknowledgements**

We thank all members of the Center for Science of Science and Innovation (CSSI) at Northwestern University, the audiences at ICSSI 2024 and ECDC 2024 for thoughtful discussions. We thank the Kellogg School of Management, the Northwestern Institute on Complex Systems (NICO), and CSSI for their funding support.

**Author Contributions**

B.W., D.W., and B.U. designed the research; B.W., and Z.M. collected data and conducted empirical analyses. All authors discussed the results and edited the manuscript.

**Competing Interest Statement**

The authors declare no competing interests.

# Supplementary Information for

# Peer Review and the Diffusion of Ideas

Binglu Wang[1,2,3], Zhengnan Ma[4], Dashun Wang[1,2,3,4], Brian Uzzi[1,3,4†]

[1] Kellogg School of Management, Northwestern University

[2] Center for Science of Science and Innovation, Northwestern University, Evanston, IL, USA

[3] Northwestern University Institute on Complex Systems, Evanston, Illinois

[4] McCormick School of Engineering, Northwestern University, Evanston, IL, USA

† *Correspondence should be addressed to B.U. ([uzzi@kellogg.northwestern.edu](uzzi@kellogg.northwestern.edu))*

## Table of Contents

# S1 Data Description

## S1.1 Peer Review Data

Our primary dataset comprises comprehensive peer review records from a single journal published by a leading medical research publisher, spanning a ten-year period from 2012 to 2022. It comprises approximately 37K manuscript submissions and 551K associated review records. Each record captures the full review invitation process, including rich manuscript metadata—such as title, abstract, author list, and corresponding author—as well as timestamped responses from reviewer invitees. The dataset also includes article-level metadata detailing editorial decisions, DOIs, and structured author rosters identifying the corresponding author for each submission.

## S1.2 Publication Data

We use SciSciNet database [1] and the Web of Science (WoS) database to extract individual publication records and citation behaviors. SciSciNet is a large-scale open-access data lake for science of science research, encompassing over 134 million scientific publications and extensive metadata on funding, public engagement, and knowledge diffusion until 2022. To track how reviewer invitees' citation behaviors evolve pre- and post-invitation, and to control for person- and paper-level covariates, we integrated six SciSciNet datasets:

- SciSciNet_Papers, which records every primary paper together with its unique ID, disciplinary category, entry counts, and foundational metrics.
- SciSciNet_Authors_Gender, which assigns each scholar a unique ID, name, and career-level indicators.
- SciSciNet_PaperAuthorAffiliations, which traces every linkage between papers, authors, and their affiliations.
- SciSciNet_PaperReferences, which captures every citing–cited pair among the primary papers.
- SciSciNet_Affiliations, which maps affiliation IDs to institution names and summary metrics.
- SciSciNet_Journals, which details journal IDs alongside titles, ISSNs, publishers, and official webpages.

- SciSciNet_PaperFields, which lists field IDs, names, and flags that indicate whether each field is top-level or a subdiscipline.

In addition, we use the WoS database to extract the citation position within each paper's reference list. This allows us to compute the prominence metric shown in Figure 4, which assesses the conceptual centrality of the corresponding author's work within the reviewer invitees' research. Specifically, we use the reciprocal of the earliest citation position as a proxy for prominence: references cited earlier in a list are assumed to be more central to the citing paper's argument.

**S1.3 Record Linkage and Panel Assembly**

We first linked invitee and author identities in the peer review dataset to unique scholar identifiers in SciSciNet. To maximize accuracy, we followed a multi-step matching procedure: (1) matching based on first initial and last name; (2) restricting candidates to SciSciNet scholars with at least three publications to ensure a robust publication record; (3) using fuzzy string matching to compare full names and affiliations, generating similarity scores for each potential match; and (4) prioritizing candidates who had previously published in the same journal or were located close to the journal in the citation network. To validate match quality, we compared corresponding authors' and reviewers' email addresses from the journal's editorial system with researchers' emails listed in publications, assuming email addresses uniquely identify individuals. This approach yielded an estimated match accuracy of over 83%.

Each manuscript was linked to the SciSciNet ID of its corresponding author, forming dyadic reviewer–submission pairs. The resulting dataset comprised 550,686 invitations tied to 37,279 unique submissions and 209,122 distinct matched reviewer invitees. Reviewer invitees' responses were then numerically encoded into three categories: Accept (clicked "Agreed"), Click-Decline (clicked "Decline" or marked "Unavailable"), and Auto-Decline (no response).

To recover local time and weekday context for each invitation, we converted the original timestamps to the reviewer invitees' local time zone—using affiliation-based location data when available or inferring time zones from the country code in the email domain, mapped to IANA standards.

This procedure yielded a longitudinal panel of 5.74 million reviewer-year observations, spanning an event-time window from 10 years prior to 5 years following the invitation. Each row in this panel includes six citation-based outcome measures (described in Figure 4), covariates, treatment indicators based on reviewer invitees' behavior, and fixed identifiers for the reviewer invitees, manuscript, and calendar year.

## S2 Descriptive Summary Between Treatment and Control Groups

In our analysis, we adjust for three categories of factors that may influence post-invitation citation behavior. First, to account for potential pre-existing ties between reviewer invitees and the manuscript's corresponding author, we construct three measures of prior relationship strength: (1) the number of co-authored papers between the reviewer invitees and the corresponding author prior to the invitation; (2) the cumulative number of times the reviewer invitees had cited the corresponding author's work; and (3) the cumulative number of times the corresponding author had cited the reviewer invitee's work. Second, to capture reviewer invitees' research experience and scholarly impact, we compute each reviewer invitees' career $h$-index (log) and total publication count (log) as of the invitation date. These measures allow us to assess whether more established researchers systematically differ in their responsiveness or exposure-related behaviors. Third, to proxy the reviewer invitees' relationship with the journal, we calculate two variables: (1) the number of publications by the reviewer invitees in the focal journal prior to the invitation, and (2) the number of times the reviewer invitees cited articles from the focal journal before the invitation. These metrics reflect prior intellectual engagement and visibility within the journal's ecosystem. These citation-based links provide a fine-grained measure of scholarly proximity that could confound post-invitation citation outcomes.

These covariates are included to ensure that any estimated differences between the Click-Decline and Auto-Decline groups are not simply attributable to underlying differences in expertise, institutional alignment, or pre-existing author-reviewer relationships.

TABLE S1: SUMMARY STATISTICS OF CONTROL AND TREATMENT GROUPS

| | Before Invitation (N = 450,737) | | | | |
|---|---|---|---|---|---|
| | Click-Decline Group (N=246,565) | | Auto-Decline Group (N=204,172) | | T-Test |
| | Mean | Std. | Mean | Std. | t-stat |
| # Collaboration for Both | 0.03 | 0.65 | 0.04 | 2.43 | -1.50 |
| # Citation Reviewer to Author | 0.57 | 5.98 | 0.55 | 8.93 | 0.62 |
| # Citation Author to Reviewer | 0.95 | 9.10 | 0.89 | 10.11 | 1.91* |
| Reviewer # Publications (log) | 4.40 | 1.23 | 4.17 | 1.36 | 58.84*** |
| Reviewer H-index (log) | 2.38 | 1.09 | 2.39 | 1.08 | -3.00*** |
| Reviewer # Pub on Journal | 1.14 | 2.78 | 0.75 | 2.25 | 51.43*** |
| Reviewer # Ref to Journal | 18.10 | 30.31 | 13.34 | 27.25 | 55.53*** |

**Table S1. Descriptive statistics of pre-invitation reviewer invitees' characteristics: Click-Decline vs. Auto-Decline groups (two-sample *t*-tests).** Variables summarized include (i) reviewer–author proximity—the number of prior co-authored papers plus cumulative citations flowing from reviewer invitees to author and vice versa; (ii) reviewer invitees' research experience—log counts of total publications and H-index; and (iii) engagement with the journal—pre-invitation publications by the reviewer invitees in the focal journal and the number of times the reviewer invitees cited articles from the focal journal. Means and standard deviations are shown for each group, with the final column reporting two-sample t-statistics. (∗∗∗ $p < 0.01$, ∗∗ $p < 0.05$, ∗ $p < 0.1$).

## S3 Supplementary Results for Difference-in-Difference Results

In the main text, we employed a canonical Difference-in-Differences (DiD) framework and documented a significant divergence in post-invitation citation behavior between Click-Decline reviewer invitees (minimally exposed) and Auto-Decline reviewer invitees (unexposed). To evaluate the robustness of these findings, we complement our primary analysis with additional specifications that address potential heterogeneity in treatment timing and unobserved individual differences, including TWFE and staggered DiD.

### S3.1 Two-Way Fixed Effect Difference-in-Differences

We estimate a static two-way fixed-effects (TWFE) difference-in-differences model that absorbs both unit-specific heterogeneity and common calendar-year shocks. Each observation is a reviewer–manuscript pair $i$ observed in year $t$:

$$y_{it} = \beta\, Treat_i \times Post_t + \gamma_i + \delta_t + \epsilon_{it}$$

where $y_{it}$ is the citation outcome for unit $i$ in period $t$; $Treat_i \times Post_t$ is the interaction term indicating treated reviewers in the post-invitation period; $\gamma_i$ is unit fixed effect; $\delta_t$ captures calendar year fixed effects respectively, and $\epsilon_{it}$ is the error term. The coefficient $\beta$ represents the difference-in-differences estimate of the invitation's effect on citations.

Estimates from this model, presented in Table S2, show results consistent with our main DiD specification. The estimated effect remains statistically significant at 0.0021 citations per year ($p = 0.000$), reflecting a 14% relative increase. Because the model includes individual fixed effects, all time-invariant covariates are absorbed and thus not separately estimated.

# TABLE S2: TWFE ESTIMATES

|  | (1) |
| --- | --- |
| $Treat_i \times Post_t$ | 0.0021*** |
|  | (0.0003) |
| Const | 0.0084*** |
|  | (0.0000) |
| Unit Fixed Effect | Y |
| Year Fixed Effect | Y |
| Observations | 4,751,559 |
| No. clusters | 446,725 |
| R-squared | 0.3877 |

**Table S2. TWFE estimates of the exposure effect via review invitation on downstream citations to corresponding authors.** This table reports estimates from our baseline TWFE difference-in-differences specification, restricted to reviewer invitees in Click-Decline and Auto-Decline groups. The dependent variable is the reviewer invitees' average annual number of citations to the focal corresponding author. Model (1) shows that reviewer invitees who *Click-Declined* an invitation cite the corresponding author 0.0021 additional times per year in the post-invitation period relative to reviewer invitees who *Auto-Declined* (p = 0.000). This estimate represents roughly a 14% increase over the post-invitation citation rate of the Auto-Decline group. The regression includes reviewer–manuscript fixed effects and calendar-year fixed effects, and excludes singleton observations, thereby controlling for all time-invariant pair characteristics and common year-specific shocks. Robust standard errors clustered at the reviewer invitee level are reported in parentheses. (*** $p < 0.01$, ** $p < 0.05$, * $p < 0.1$).

To ensure the robustness of our event-study results, we update the TWFE specification by replacing the single $Treat_i \times Post_t$ term with interactions between treatment status and a set of event-time dummies. The results, shown in Figure S1, confirm the parallel trends assumption and are consistent with our main findings.

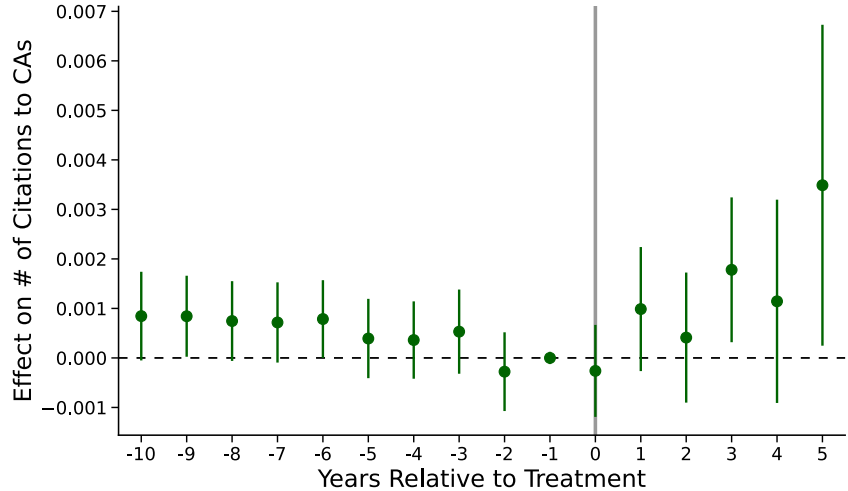

**Figure S1. Event-study estimates from dynamic TWFE.** Estimates include relative-year indicators interacted with treatment status, un-interacted event-time dummies, and the same fixed effects used in the static TWFE model. Each marker plots $\beta_k$ from the event-study specification with fixed effects, measuring the difference in the average yearly number of citations a reviewer invitee gives the focal corresponding author (CA) $k$ years relative to the invitation with one year before omitted as the baseline. Bars show the 95% confidence interval derived from the regression standard errors. Pre-treatment coefficients hover around zero. Post-invitation effects turn positive and grow steadily, reaching roughly 0.003 additional citations per year five years later.

### S3.2 Staggered Difference-in-Differences

In our setting, reviewer invitees are exposed to treatment at different points in time, and the duration of post-treatment observation varies across individuals. For estimates from both the canonical Difference-in-Differences and traditional two-way fixed effects models to be interpreted as valid average treatment effects, they require the assumption of homogeneous treatment effects across cohorts and over time. However, as demonstrated by Goodman-Bacon [2], TWFE estimators aggregate all possible two-by-two comparisons—including comparisons between early- and late-treated groups—which can introduce negative weights and yield biased estimates in the presence of treatment effect heterogeneity.

To address this concern, we implement the CSDiD estimator proposed by Callaway and Sant'Anna [3]. This method improves *ATT* (i.e., average treatment effect on the treated) estimation under staggered treatment timing by estimating group-time average treatment effects and appropriately weighting them. Unlike TWFE, CSDiD avoids the problematic implicit weighting structure and produces unbiased estimates even when treatment effects vary across cohorts or time. It estimates a cohort- and time-specific average treatment effect:

$$ATT(g,t) = E[Y_{it}(1) - \widetilde{Y_{it}}(0) \mid G_i = g]$$

where $g$ denotes the year when a unit is first treated, $t$ is the calendar year, $Y_{it}(1)$ is the observed outcome under treatment, and $\widetilde{Y_{it}}(0)$ is the counterfactual outcome had the unit remained untreated. This effect is defined only for post-treatment periods ($t \geq g$). We estimate $ATT(g,t)$ with *CSDiD*, which delivers three ways to collapse the cell-level effects which are reported in Table S3:

- Simple: Each cohort-time cell's $ATT(g,t)$ is weighted by the number of observations in that cell, and the overall effect is the sample-size-weighted mean across all $g$ and $t$.
- Cohort-weighted: Within every treated cohort $g$, first take a sample-size-weighted average of its $ATT(g,t)$ across post-treatment periods; then weight those cohort-level means by the total number of observations in each cohort.
- Calendar-weighted: For each post-treatment calendar year $t$, average $ATT(g,t)$ across cohorts using sample-size weights; then combine the year-specific means by weighting them according to the number of treated observations appearing in that year.

By estimating effects cohort-by-cohort and then averaging, the *CSDiD* estimator sidesteps the contamination bias that plagues conventional two-way fixed-effects models under staggered adoption and permits flexible reporting of dynamic or cohort-specific treatment patterns.

Model (1) of Table S3 applies the simple Callaway–Sant'Anna aggregation and yields an *ATT* of $0.0011 (p = 0.000)$, which reflects that Click-Decline reviewer invitees cite corresponding authors 0.0011 more times per year than Auto-Decline reviewer invitees after the invitation. While modest in absolute terms, this still reflects a 7% relative increase compared to the Auto-Decline group's post-invitation citation rate. The cohort-weighted (Model 2) and calendar-weighted (Model 3) specifications similarly confirm a significant and consistent positive effect. Additionally, the event study estimates in Figure S2 show that the increase begins in the year of the invitation and persists over time.

TABLE S3: STAGGERED DID ESTIMATES

|  | (1) | (2) | (3) |
|---|---|---|---|
| ATT | 0.0011*** | 0.0007** | 0.0036* |
|  | (0.0004) | (0.0003) | (0.0019) |
| Observations | 4,656,924 | 4,656,924 | 4,656,924 |
| Type | simple | cohort | calendar |

**Table S3. Average treatment effects from staggered DiD specifications.** This table reports the average treatment effect on the treated (ATT) estimated with the Callaway-Sant'Anna staggered difference-in-differences approach under three aggregation schemes: simple (Model 1), cohort-weighted (Model 2), and calendar-weighted (Model 3). Across all three aggregation schemes—simple (0.0011, $p = 0.002$), cohort-weighted (0.0007, $p = 0.015$), and calendar-weighted (0.0036, $p = 0.061$)—the ATT remains positive and statistically significant, indicating a consistently robust treatment effect. Robust standard errors clustered at the reviewer invitee level are shown in parentheses. (*** $p < 0.01$, ** $p < 0.05$, * $p < 0.1$).

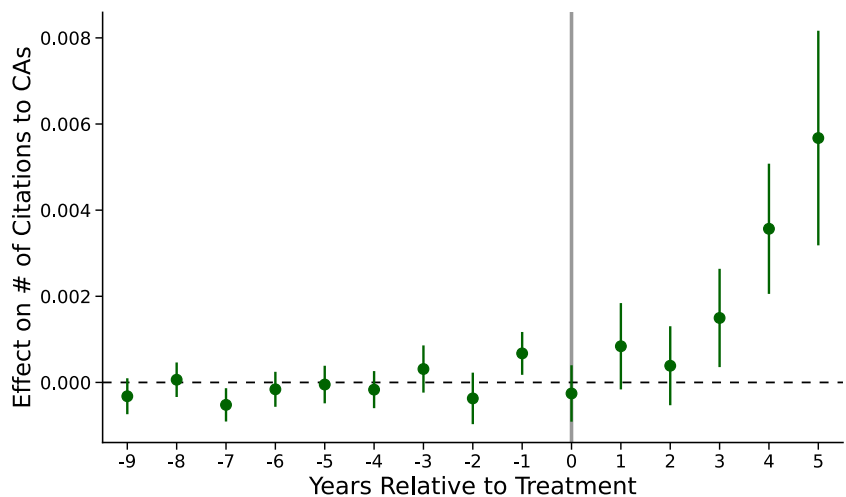

**Figure S2. Staggered DiD event-study coefficients.** The plot shows event-specific *ATT* estimates from the Callaway–Sant'Anna estimator, first averaged within each relative year *t* and then aggregated across cohorts with cohort-size weights. Each marker represents the estimated change in a reviewer invitee's yearly citations to the focal corresponding author (CA) *k* years relative to the invitation. Whiskers depict 95% confidence intervals based on the *CSDiD* robust standard errors. Coefficients for pre-treatment years fluctuate around zero, supporting parallel trends, whereas post-invitation effects become increasingly positive, reaching roughly 0.006 additional citations per year five years after treatment. Note that $k = -10$ is not reported because *CSDiD* forms each pre-treatment contrast relative to the immediately prior year, and the window stops at –10.

## S4 Supplementary Results for Instrumental Variable Estimation

We adopt a two-stage least-squares (2SLS) approach for instrumental variable estimation. The instrument is a binary indicator for whether the invitation was sent on a weekend, and the unit of analysis remains the reviewer–manuscript pair. Beginning with the same panel used in our DiD estimations, we collapse each reviewer–manuscript trajectory to two observations—the mean outcome over the entire pre-invitation window and the mean outcome over the post-invitation window—before running the IV procedure.

In the first-stage regressions, the weekend-assignment instrument is exceptionally strong: the Kleibergen–Paap rk F-statistic is 1,275.76 in the model without control variables and 1,220.48 in the specification that includes control variables—both far exceeding the Stock & Yogo's [4] 10% weak-instrument critical value of 16.38, safeguarding the 2SLS estimates against weak-instrument bias.

## TABLE S4: FIRST-STAGE RESULTS OF IV TEST

|  | (1) | (2) |
|---|---|---|
| **IV_weekend** | **0.0469*** | **0.0457*** |
|  | **(0.0013)** | **(0.0013)** |
| # Collaboration for Both |  | -0.0004* |
|  |  | (0.0002) |
| # Citation Reviewer to Author |  | 0.0003*** |
|  |  | (0.0001) |
| # Citation Author to Reviewer |  | -0.0002*** |
|  |  | (0.0001) |
| Reviewer # Publications (log) |  | 0.0077*** |
|  |  | (0.0004) |
| Reviewer H-index (log) |  | -0.0022*** |
|  |  | (0.0004) |
| Reviewer # Pub on Journal |  | -0.0032*** |
|  |  | (0.0002) |
| Reviewer # Ref to Journal |  | 0.0008*** |
|  |  | (0.0000) |
| Const | 0.1837*** | 0.1463*** |
|  | (0.0012) | (0.0021) |
| Kleibergen–Paap rk F-statistics | 1275.76 | 1220.48 |
| Observations | 742,229 | 742,229 |
| No. clusters | 450,773 | 450,773 |
| R-squared | 0.0010 | 0.0053 |

**Table S4. First-stage regression from the 2SLS instrumented analysis.** This table reports first-stage estimates from a two-stage least squares (2SLS) regression model in which the binary instrument *IV_weekend* indicates whether the invitation was sent on a weekend. Model (2) controls for prior collaboration with the corresponding author, reviewer invitees' research experience, and ties to the journal. The Kleibergen–Paap rk

F-statistics of 1,275.76 in Model (1) and 1,220.48 in Model (2) dwarf the Stock–Yogo (2005) 10% critical value of 16.38, confirming the instrument's strength in both specifications. Robust standard errors clustered at the reviewer invitee level are shown in parentheses. (∗∗∗ $p < 0.01$, ∗∗ $p < 0.05$, ∗ $p < 0.1$).

## S5 Supplementary Results for Alternative Outcome Variables

For every reviewer invitee, we summarized annual citing behavior along six dimensions. Direct Attention captures the proportion of that year's papers in which the reviewer invitees cite the invited article itself. Breadth records the proportion that cite any work by the manuscript's corresponding author. When those citations occur, we measure their intensity: Depth is the mean count of such citations per reviewer invitee's paper, and Normalized Depth rescales the same counts by each paper's reference-list length before averaging. To characterize variety, we calculate Diversity, the number of distinct corresponding author papers cited in that year. Finally, we quantify Prominence by identifying the earliest position at which the author's work appears in a paper's reference list, taking the reciprocal of this position as the measure. For cases where no citation occurs, we assign a value of zero. Earlier positions imply greater importance or centrality to the paper. All six metrics are computed for every calendar year from ten years before to five years after each invitation, using publication and reference data drawn from SciSciNet and WoS described above.

Table S5 presents an overview of the outcome measures employed in the analysis, detailing their conceptual motivation and formal definitions.

# TABLE S5: SUMMARY OF OUTCOME VARIABLES

| Outcome Variable | Explanation | Definition |
|---|---|---|
| **Direct Attention** | Measures whether researchers later cite the specific manuscript they were invited to review, conditional on that manuscript being published. Indicates direct influence from the exposure. | Proportion of reviewer invitees' publications each year that cite the invited manuscript: $\frac{N_{pubs\_citing\_manuscript}}{N_{pubs\_per\_year}}$ |
| **Breadth** | Assesses the extent to which reviewer invitees cite any work by the corresponding author (beyond the focal manuscript), capturing broader engagement with the author's research. | Proportion of reviewer invitees' publications each year that cite any paper by the corresponding author: $\frac{N_{pub\_citing\_author}}{N_{pubs\_per\_year}}$ |
| **Depth (Figure 2)** | Evaluates how extensively a reviewer invitee cites the corresponding author's work in each publication. Higher values reflect more intensive engagement. | Average number of references to the corresponding author each year: $\frac{1}{N_{pubs}}\sum_{i=1}^{N_{pubs}} \#refs_i^{to\_author}$ |
| **Normalized Depth** | Adjusts depth by accounting for total number of references in each publication. Captures relative intensity of citations to the author. | Average proportion of references per publication attributed to the corresponding author each year: $\frac{1}{N_{pubs}}\sum_{i=1}^{N_{pubs}} \left(\frac{\#refs_i^{to\_author}}{\#refs_i}\right)$ |
| **Diversity** | Captures the range of the corresponding author's work cited by the reviewer invitees. Higher values indicate engagement with multiple distinct papers, not just repeated citation of one. | Total number of unique papers by the corresponding author cited each year: $N_{unique\_cited\_papers\_by\_author}$ |
| **Prominence** | Indicates the prominence of the corresponding author's work in the reviewer invitee's reference list, based on citation position. Earlier positions imply greater importance or centrality to the paper. | The reciprocal of earliest reference list position of the author's work across reviewer invitees' publications each year: $\frac{1}{\sum_{i=1}^{N_{pubs}} min\left(Ref\_Position_{author,i}\right)}$ |

**Table S5. Definitions of outcome variables capturing downstream citation behavior.** The table lists each outcome variable's conceptual focus—ranging from direct attention to the invited manuscript to breadth, depth, diversity, and prominence of citations to the corresponding author—and provides its exact operational formula.

We compare reviewer invitees who click-declined versus those who auto-declined the invitation across six citation-based metrics mentioned above using two-sample t-tests for both post-invitation (Figure 4) and pre-invitation (Figure S3) periods. For pre-invitation, the two groups do not differ significantly in direct attention, diversity, or prominence (Figure S3AEF), but exhibit significant difference in breadth (Figure S3B, $p = 0.001$) and depth (Figure S3C, $p = 0.001$), and a marginal difference in normalized depth (Figure S3D, $p = 0.077$). Although some dimensions show statistically significant pre-invitation differences, they consistently run in the opposite direction of our primary post-invitation findings. This pattern suggests that our observed post-invitation differences likely represent conservative estimates of the peer review exposure effect.

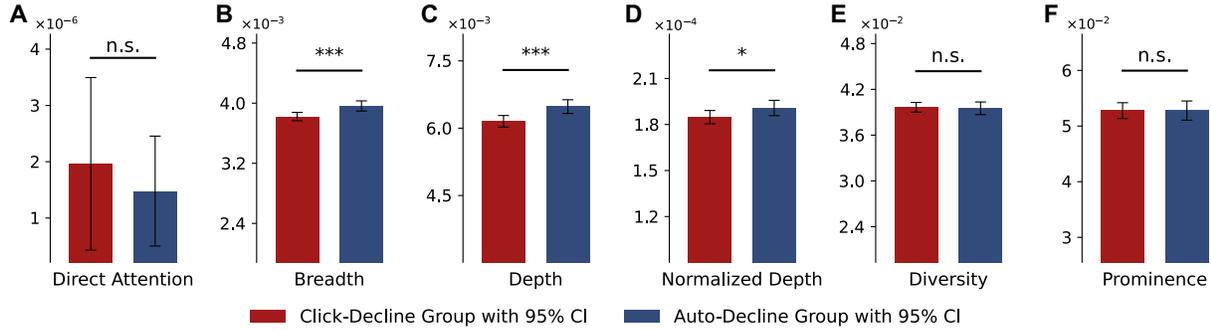

**Figure S3. Pre-invitation comparisons across six dimensions.** We compare the Click-Decline and Auto-Decline groups using two-sample t-tests on six citation-based outcome measures, for pre-invitation periods. Group means are displayed with 95% confidence intervals, with significance levels and conceptual groupings annotated. Outcome variables include: (A) Direct Attention; (B) Breadth; (C) Depth; (D) Normalized Depth; (E) Diversity; and (F) Prominence. Color coding indicates conceptual clusters. Pre-invitation differences between Click-Decline and Auto-Decline groups were examined across the six dimensions of citation behavior. While several dimensions show statistically significant pre-invitation difference (Breadth, $p = 0.001$; Depth, $p = 0.001$; and Normalized Depth, $p = 0.077$), these differences run notably in directions opposite to our primary findings. This suggests our post-invitation differences represent conservative estimates of the peer review effect. (*** $p < 0.01$, ** $p < 0.05$, * $p < 0.1$).

The DiD specification in Table S6 is identical to the baseline model reported in the main text, with the only difference being the dependent variable:

$$y_{it} = \beta_0\, Treat_i \times Post_t + \beta_1\, Treat_i + \beta_2\, Post_t + \beta_3\, C_i + \epsilon_{it}$$

Across all six measures—Direct Attention, Breadth, Depth, Normalized Depth, Diversity, and Prominence—the point estimates are positive and statistically significant ($p < 0.01$). The results indicate that exposure via the review invitation not only boosts the likelihood of citing the specific manuscript but also broadens, deepens, diversifies, and elevates the author's presence in reviewer invitees' subsequent research.

TABLE S6: ALTERNATIVE OUTCOME VARIABLE ESTIMATES

|  | Direct Attention | Breadth | Depth | Normalized Depth | Diversity | Prominence |
|---|---|---|---|---|---|---|
| $Treat_i \times Post_t$ | 0.0001*** | 0.0012*** | 0.0021*** | 0.0001*** | 0.0259*** | 0.0377*** |
|  | (0.0000) | (0.0002) | (0.0004) | (0.0000) | (0.0028) | (0.0056) |
| $Treat_i$ | 0.0000 | -0.0001* | -0.0003* | 0.0000 | 0.0001 | -0.0000 |
|  | (0.0000) | (0.0001) | (0.0002) | (0.0000) | (0.0012) | (0.0017) |
| $Post_t$ | 0.0004*** | 0.0050*** | 0.0087*** | 0.0002*** | 0.0726*** | 0.1352*** |
|  | (0.0000) | (0.0001) | (0.0003) | (0.0000) | (0.0019) | (0.0040) |
| Const | 0.0000*** | 0.0040*** | 0.0065*** | 0.0002*** | 0.0395*** | 0.0523*** |
|  | (0.0000) | (0.0001) | (0.0001) | (0.0000) | (0.0010) | (0.0013) |
| Observations | 4,755,571 | 4,755,571 | 4,755,571 | 4,755,571 | 4,755,571 | 4,755,571 |
| No. clusters | 450,737 | 450,737 | 450,737 | 450,737 | 450,737 | 450,737 |
| R-squared | 0.0009 | 0.0026 | 0.0014 | 0.0010 | 0.0035 | 0.0019 |

**Table S6. The effect of exposure via review invitation on six dimensions.** The table reports the Difference-in-Differences estimates across six dimensions. We run DiD regressions restricted to Click-Decline and Auto-Decline on six citation-based outcome measures. The treatment effect is positive and highly significant for every outcome variable: Direct Attention = 0.0001 ($p$ = 0.007), Breadth = 0.0012 ($p$ = 0.000), Depth = 0.0021 ($p$ = 0.000), Normalized Depth = 0.0001 ($p$ = 0.000), Diversity = 0.0259 ($p$ = 0.000), and Prominence = 0.0377 ($p$ = 0.000). These coefficients correspond to relative increases of 25%, 13%, 14%, 25%, 23% and 20% in Direct Attention, Breadth, Depth, Normalized Depth, Diversity, and Prominence, respectively, compared with the Auto-Decline group's post-invitation citation rates. Robust standard errors clustered at the reviewer invitee level are shown in parentheses. (*** $p$ < 0.01, ** $p$ < 0.05, * $p$ < 0.1).